\documentclass[aps,prl,twocolumn,showpacs,superscriptaddress,groupedaddress]{revtex4-1} 

\usepackage{graphicx}        
\usepackage{dcolumn}        
\usepackage{bm}             
\usepackage{amssymb}
\usepackage{amsmath}        
\usepackage{color}
\definecolor{korr_26Apr}{rgb}{0,0,0} 
\definecolor{red}{rgb}{1,0,0}

\def \d{\mathrm{d}}

\begin{document}

\widetext

\title{Flux saturation length of sediment transport}
\author{Thomas P\"ahtz$^{1,2}$, Jasper F. Kok$^3$, Eric J. R. Parteli$^4$ and Hans J. Herrmann$^{5,6}$}
\affiliation{1.~Department of Ocean Science and Engineering, Zhejiang University, 310058 Hangzhou, China. \\
2.~State Key Laboratory of Satellite Ocean Environment Dynamics, Second Institute of Oceanography, 310012 Hangzhou, China. \\
3.~${\mbox{Department of Earth and Atmospheric Sciences, Cornell University, Ithaca, NY 14850, USA.}}$ \\
4.~${\mbox{Institute for Multiscale Simulation, Universit\"at Erlangen-N\"urnberg, N\"agelsbachstra{\ss}e~49b, 91052 Erlangen, Germany.}}$ \\
5.~Departamento de F\'isica, Universidade Federal do Cear\'a, 60451-970 Fortaleza, Cear\'a, Brazil. \\
6.~Computational Physics, IfB, ETH Z\"urich, Schafmattstra{\ss}e~6, 8093 Z\"urich, Switzerland.}

%\date{\today}

\begin{abstract}
Sediment transport along the surface drives geophysical phenomena as diverse as wind erosion and dune formation. The main length-scale controlling the dynamics of sediment erosion and deposition is the saturation length $L_{\mathrm{s}}$, which characterizes the flux response to a change in transport conditions. Here we derive, for the first time, an expression predicting $L_{\mathrm{s}}$ as a function of the average sediment velocity under different physical environments. Our expression accounts for both the characteristics of sediment entrainment and the saturation of particle and fluid velocities, and has only two physical parameters which can be estimated directly from independent experiments. We show that our expression is consistent with measurements of $L_{\mathrm{s}}$ in both aeolian and subaqueous transport regimes over at least five orders of magnitude in the ratio of fluid and particle density, including on Mars. 

\end{abstract}
\pacs{45.70.-n, 47.55.Kf, 92.40.Gc}

\maketitle
Sediment transport along the surface drives a wide variety of geophysical phenomena, including wind erosion, dust aerosol emission, and the formation of dunes and ripples on ocean floors, river beds, and planetary surfaces \cite{Bagnold_1941,van_Rijn_1993,Garcia_2007,Shao_2008,Bourke_et_al_2010,Kok_et_al_2012}. The primary transport modes are {\em{saltation}}, which consists of particles jumping downstream close to the ground at nearly ballistic trajectories, and {\em{creep}} (grains rolling and sliding along the surface). A critical parameter in sediment transport is the distance needed for the particle flux to adapt to a change in flow conditions, which is characterized by the {\em{saturation length}}, $L_{\mathrm{s}}$. 
%In spite of insights gained from recent studies \cite{Andreotti_et_al_2010,Franklin_and_Charru_2011,Ma_and_Zheng_2011}, 
Predicting $L_{\mathrm{s}}$ under given transport conditions remains a long-standing open problem \cite{Kok_et_al_2012,Andreotti_et_al_2010,Franklin_and_Charru_2011,Ma_and_Zheng_2011,Sauermann_et_al_2001}. 

Indeed, $L_{\mathrm{s}}$ partially determines the dynamics of 
%bedforms such as dunes, 
dunes,
for instance by dictating the wavelength of the smallest (``elementary'') dunes on a sediment surface \cite{Kroy_et_al_2002,Fourriere_et_al_2010} and the minimal size of crescent-shaped 
%barchan dunes 
barchans \cite{Kroy_et_al_2002,Parteli_et_al_2007}. Moreover, although flux saturation plays a significant role for the evolution of fluvial sediment landscapes \cite{Cao_et_al_2012}, morphodynamic models used in hydraulic engineering usually treat $L_{\mathrm{s}}$ as an adjustable parameter \cite{He_et_al_2009}. The availability of an accurate theoretical expression predicting $L_{\mathrm{s}}$ for given transport conditions would thus be an important contribution to the planetary, geological and engineering sciences. In this Letter, we present such a theoretical expression for $L_{\mathrm{s}}$. In contrast to previously proposed relations for $L_{\mathrm{s}}$, the expression presented here explicitly accounts for the relevant forces that control the relaxation of particle and fluid velocities, and also incorporates the distinct entrainment mechanisms prevailing in aeolian and subaqueous transport (defined below).

The average momentum of transported grains per unit soil area, the sediment flux $Q$, is defined as $Q=MV$, where $M$ is the mass of sediment in flow per unit soil area, and $V$ is the average particle velocity. Since the fluid loses momentum to accelerate the particles, $Q$ is limited by a steady-state value, the saturated flux $Q_{\mathrm{s}}$. This flux is largely set by the fluid density ${\rho}_{\mathrm{f}}$ and the fluid shear velocity $u_{\ast}$ \cite{Bagnold_1941,van_Rijn_1993,Garcia_2007,Shao_2008,Kok_et_al_2012}, which is proportional to the mean flow velocity gradient in turbulent boundary layer flow \cite{Kok_et_al_2012}. In typical situations, such as on the streamward side of dunes, the deviation of $Q$ from $Q_{\mathrm{s}}$ is small, that is, $|1-Q/Q_{\mathrm{s}}|\ll1$ \cite{Kroy_et_al_2002,Sauermann_et_al_2001,Duc_and_Rodi_2008}. The rate $\Gamma(Q)$ of the relaxation of $Q$ towards $Q_{\mathrm{s}}$ in the downstream direction ($x$) can thus be approximately written as \cite{Sauermann_et_al_2001,Duc_and_Rodi_2008,Andreotti_et_al_2010},
\begin{eqnarray}
%\Gamma(Q)=\frac{\d Q}{\d x}&\approxeq&\frac{Q_{\mathrm{s}}-Q}{L_{\mathrm{s}}}, \label{Ls}
\Gamma(Q)= {\d Q}/{\d x}&\approxeq& {\left[{Q_{\mathrm{s}}-Q}\right]}/{L_{\mathrm{s}}}, \label{Ls}
\end{eqnarray}
where $\Gamma$ is Taylor-expanded to first order around $Q=Q_{\mathrm{s}}$ ($\Gamma(Q_{\mathrm{s}})=0$), and the negative inverse Taylor-coefficient gives the saturation length, $L_{\mathrm{s}}$. Flux saturation is controlled by the downstream evolutions of $M$ and $V$ towards their respective steady-state values, $M_{\mathrm{s}}$ and $V_{\mathrm{s}}$. Changes in $M$ with $x$ are controlled by particle entrainment from the sediment bed into the transport layer. In the aeolian regime (dilute fluid such as air), entrainment occurs predominantly through particle impacts \cite{Kok_et_al_2012}, whereas in the subaqueous regime (dense fluid such as water) entrainment occurs mainly through fluid lifting \cite{van_Rijn_1993,Garcia_2007}. On the other hand, the evolution of $V$ towards $V_{\mathrm{s}}$ is mainly controlled by the acceleration of the particles due to fluid drag, and their deceleration due to grain-bed collisions \cite{Sauermann_et_al_2001,Fourriere_et_al_2010}. We note that the evolution of $V$ is affected by changes in $M$ and vice versa. For instance, an increase of $M$ leads to a decrease in $V$ in the absence of horizontal forces due to conservation of horizontal momentum. For simplicity, previous studies neglected either the saturation of $V$ \cite{Sauermann_et_al_2001,Charru_2006} or the relaxation of $M$, as well as changes in $V$ due to grain-bed collisions \cite{Andreotti_et_al_2010,Fourriere_et_al_2010}. Moreover, all previous studies 
%\cite{Sauermann_et_al_2001,Charru_2006,Andreotti_et_al_2010,Fourriere_et_al_2010} 
did not account for the relaxation of the fluid velocity ($U$) towards its steady-state value ($U_{\mathrm{s}}$) within the transport layer. This relaxation is driven by changes in the transport-flow feedback resulting from the relaxations of $M$ and $V$. For instance, increasing $V$ reduces the relative velocity $V_{\mathrm{r}} = U - V$ and thus the fluid drag. In turn, as $V_{\mathrm{r}}$ decreases, the amount of momentum transferred from the fluid to the transport layer also decreases, which results in an increase in $U$, whereas an increase in $U$ again increases $V_{\mathrm{r}}$.

In this Letter, we derive a theoretical expression for $L_{\mathrm{s}}$ which encodes {\em{all}} aforementioned relaxation mechanisms.
% and is consistent with measurements over at least 5 orders of magnitude in the ratio of the fluid and particle density, without requiring fitting to any of these measurements. 
Indeed, since previously proposed relations for $L_{\mathrm{s}}$ neglect some of the interactions that determine $L_{\mathrm{s}}$ \cite{Sauermann_et_al_2001,Kroy_et_al_2002,Andreotti_et_al_2010,Fourriere_et_al_2010}, it is uncertain how to adapt these equations to compute $L_{\mathrm{s}}$ in extraterrestrial environments, such as Mars \cite{Parteli_et_al_2007,Bourke_et_al_2010,Kok_et_al_2012}. Our theoretical expression overcomes this problem, since it is valid for arbitrary physical environments for which turbulent fluctuations of the fluid velocity, and thus transport as suspended load \cite{Kok_et_al_2012}, can be neglected. For aeolian transport under terrestrial conditions, this regime corresponds to $u_{\ast} \lesssim 4u_{\mathrm{t}}$, where $u_{\mathrm{t}}$ is the threshold $u_{\ast}$ for sustained transport \cite{van_Rijn_1993,Garcia_2007,Kok_et_al_2012}. 

%The first step towards obtaining an expression for $L_{\mathrm{s}}$ is to obtain the momentum conservation equation describing the momentum change of the transport layer. In order to obtain an expression for $L_{\mathrm{s}}$, we 
We start from the momentum conservation equation for steady ($\partial/\partial t=0$) dilute granular flows \cite{Jenkins_and_Richman_1985},
\begin{eqnarray}
%\frac{\partial\rho\langle v_x^2\rangle}{\partial x}+\frac{\partial\rho\langle v_xv_z\rangle}{\partial z}=\langle f_x\rangle, \label{mom1}
{\partial\rho\langle v_x^2\rangle}/{\partial x} + {\partial\rho\langle v_xv_z\rangle}/{\partial z}=\langle f_x\rangle, \label{mom1}
\end{eqnarray}
where $\langle\rangle$ denotes the ensemble average, $\rho$ the mass density, $\mathbf{v}$ the particle velocity, and $\mathbf{f}$ the external body force per unit volume applied on a sediment particle. Here $\mathbf{f}$ incorporates the main external forces acting on the transported particles: drag, gravity, buoyancy, and added mass. The added mass force arises because the speed of the fluid layer immediately surrounding the particle is closely coupled to that of the particle, thereby enhancing the particle's inertia by a factor $1+0.5s^{-1}$, where $s\!=\!\rho_{\mathrm{p}}/\rho_{\mathrm{f}}$ is the grain-fluid density ratio \cite{van_Rijn_1993}. Although this added mass effect is negligible in aeolian transport ($0.5\,s^{-1} \ll 1$), it affects the motion of particles in the subaqueous regime  \cite{van_Rijn_1993}. Integration of Eq.~(\ref{mom1}) over the entire transport layer depth ($\int_0^\infty..\d z$) yields,
\begin{eqnarray}
 \frac{\d(c_{v}MV^2)}{\d x}=\int\limits_0^\infty\langle f_x\rangle\d z+(\rho\langle v_xv_z\rangle)(0), \label{mom2}
\end{eqnarray}
where $M=\int_0^\infty\rho\d z$, $V=\int_0^\infty\rho\langle v_x\rangle\d z/M$, and $c_v=\int_0^\infty\rho\langle v_x^2\rangle\d z/(MV^2)$. In Eq.~(\ref{mom2}), the quantity $(\rho\langle v_xv_z\rangle)(0)$ gives the difference between the average horizontal momentum of particles impacting onto ($-(\rho_\downarrow\langle v_xv_z\rangle_\downarrow)(0)$) and leaving ($(\rho_\uparrow\langle v_xv_z\rangle_\uparrow)(0)$) the sediment bed per unit time and soil area. This momentum change is consequence of the collisions between particles within the sediment bed ($z\leq 0$). 
Thus, $(\rho\langle v_xv_z\rangle)(0)$ is an effective frictional force which the soil applies on the transport layer per unit soil area. It is proportional to the normal component of the force which the transport layer exerts onto the sediment bed \cite{Garcia_2007,Sauermann_et_al_2001,Paehtz_et_al_2012}, $(\rho\langle v_xv_z\rangle)(0)=-\mu gM(s-1)/(s+0.5)$, where $\mu$ is the associated Coulomb friction coefficient, and $g$ the gravitational constant. 
In order to obtain the momentum conservation equation of the particles within the transport layer from Eq.~(\ref{mom2}), we first note that $\int\limits_0^\infty\langle f_x^{\mathrm{drag}}\rangle\d z\approx\frac{3M}{4sd}\cdot C_d(V_{\mathrm{r}})\cdot V_{\mathrm{r}}^2$ \cite{Paehtz_et_al_2012}, where $d$ is the mean grain diameter, while $C_d(V_\mathrm{r})$ is the drag coefficient associated with the fluid drag on transported particles, which is intermediate to fully viscous drag ($C_d\propto\nu/[V_\mathrm{r}d]$, with $\nu$ standing for the kinematic viscosity) and fully turbulent drag (constant $C_d$).
%and $V_{\mathrm{r}}=U-V$ is the difference between the average fluid ($U$) and particle velocity ($V$).
By further noting that the change of $c_v$ with $x$ is negligible 
(see Suppl.~Mat.~\cite{Suppl_Mat}), we obtain,
\begin{equation}
 c_{v}\frac{\d(MV^2)}{\d x}=\frac{3M}{4(s+0.5)d}\cdot C_d(V_{\mathrm{r}})\cdot V_{\mathrm{r}}^2-\frac{s-1}{s+0.5}\mu gM. \label{momconserv}
 %c_{v}\d(MV^2)/\d x=3M/(4c_asd)\cdot C_d(V_{\mathrm{r}})\cdot V_{\mathrm{r}}^2-\mu\tilde gM/c_a. \label{momconserv}
\end{equation}
Next, we solve Eq.~(\ref{momconserv}) for $\frac{\d V}{\d x}$ thus obtaining an equation of the form $\frac{\d V}{\d x} = \Omega(V)$, and we expand $\Omega(V)$ around saturation, that is, $\Omega(V)\approx\Omega(V_{\mathrm{s}})+(V\!-\!V_{\mathrm{s}})\,{\d \Omega/\d V|_{V_{\mathrm{s}}}}$. By noting that $\Gamma(V)=\frac{\d Q}{\d x}(V)=\left(M(V)+V\frac{\d M(V)}{\d V}\right)\Omega(V)$ and $\Omega(V_{\mathrm{s}})=0$, we obtain $L_{\mathrm{s}} = -(\d \Gamma / \d Q)^{-1}|_{Q = Q_{\mathrm{s}}} = -(\d \Omega / \d V)^{-1}|_{V = V_{\mathrm{s}}}$, which leads to,
\begin{equation}
L_{\mathrm{s}}=(s+0.5)c_v{(2+c_M)V_{\mathrm{s}}}V_{\mathrm{rs}}\mathrm{F}\mathrm{K}\cdot{[\mu(s-1)g]}^{-1}, \label{LVfinal2} 
\end{equation}
where $c_M=\frac{V_{\mathrm{s}}}{M_{\mathrm{s}}}\frac{\d M}{\d V}(V_{\mathrm{s}})$, and $K=\left(1-\frac{\d U}{\d V}(V_{\mathrm{s}})\right)^{-1}$, while $V_{\mathrm{rs}}$ (the steady-state value of $V_{\mathrm{r}}$) and $F$ are given by, 
\begin{eqnarray}
&&V_{\mathrm{rs}}=\left[{{\sqrt{8\mu(s-1)gd/9 +({8\nu}/d)^2}-8\nu/d}}\right], \ \ \ {\mbox{and,}} \label{vrs3} \\
&&\mathrm{F}={\left[{{V_{\mathrm{rs}}+16\nu/d}}\right]}\cdot{\left[{{2V_{\mathrm{rs}}+16\nu/d}}\right]}^{-1}, \label{F}
\end{eqnarray}
respectively. Eqs.~(\ref{vrs3}) and (\ref{F}) result from using $C_d(V_{\mathrm{r}})=\frac{24{\nu}}{{V_{\mathrm{r}}}d} + 1.5$ (valid for natural sediment \cite{Julien_1995}). We find that using other reported drag laws only marginally affects the value of $L_{\mathrm{s}}$. 
Furthermore, we note that in the subaqueous regime $c_M \approx 0$, since in this regime $M$ changes within a time-scale which is more than one order of magnitude larger than the time-scale over which $Q$ changes \cite{Duran_et_al_2012}. This difference in time-scales implies $V\d M\ll\d Q$ and thus $V\d M\ll M\d V$ in the subaqueous regime. In contrast, in the aeolian regime, $c_M\approx 1$ as the total mass of ejected grains upon grain-bed collisions is approximately proportional to the speed of impacting grains \cite{Kok_and_Renno_2009}, which yields $M/M_{\mathrm{s}}\approx V/V_{\mathrm{s}}$.

In Eq.~(\ref{LVfinal2}), the quantity $K$ encodes the effect of the relaxation of the transport-flow feedback, neglected in previous works \cite{Sauermann_et_al_2001,Andreotti_et_al_2010,Charru_2006}. In the subaqueous regime, this transport-flow feedback has a negligible influence on the fluid speed \cite{Duran_et_al_2012} (and thus on its relaxation). In this regime, $\frac{\d U}{\d V}(V_{\mathrm{s}})\approx0$, which yields $K\approx1$ and thus,
\begin{eqnarray}
 L_{\mathrm{s}}^{\mathrm{subaq}}&=&[2s+1]c_v{V_{\mathrm{s}}}V_{\mathrm{rs}}\mathrm{F}\cdot{[\mu(s-1)g]}^{-1}. \label{LVfinal2subq}
\end{eqnarray}
In contrast, in the aeolian regime, $U$ scales with the shear velocity at the bed ($u_\mathrm{b}$) \cite{Paehtz_et_al_2012,Duran_et_al_2012}, and thus ${\frac{\d U}{\d V}}(V_{\mathrm{s}}) \approx \frac{U_{\mathrm{s}}}{u_{\mathrm{bs}}}\frac{\d u_\mathrm{b}}{\d V}(V_{\mathrm{s}})$, where $u_\mathrm{bs}$ is the steady-state value of $u_\mathrm{b}$. Using the mixing length approximation of inner turbulent boundary layer equations \cite{George_2009}, $u_\mathrm{b}$ can be expressed as $u_\mathrm{b} = u_{\ast}{\left[{1-3MC_d(V_{\mathrm{r}})V_{\mathrm{r}}^2/(4(s+0.5)d{\rho}_{\mathrm{f}}u_{\ast}^2)}\right]}^{1/2}$ \cite{Duran_et_al_2012}. By using this expression to compute $\frac{\d u_\mathrm{b}}{\d V}$ and noting that $u_\mathrm{bs} \approx u_{\mathrm{t}}$ \cite{Kok_et_al_2012}, we obtain the following expression for $K$, 
\begin{eqnarray}
\mathrm{K}=\frac{1+{{{\mathrm{F}}}^{-1}{\left[{(V_{\mathrm{s}}\!+\!V_{\mathrm{rs}})/(2V_{\mathrm{rs}})}\right]}}\cdot[{\left({{u_{\ast}}/{u_{\mathrm{t}}}}\right)}^2\!-\!1]}{1+{{\left[{(V_{\mathrm{s}}+V_{\mathrm{rs}})/(2V_{\mathrm{s}})}\right]}}\cdot[{\left({{u_{\ast}}/{u_{\mathrm{t}}}}\right)}^2\!-\!1]}. \label{Lsfluid}
\end{eqnarray}
Using Eq.~(\ref{Lsfluid}) to compute $K$, 
$L_{\mathrm{s}}$
%the saturation length 
in the aeolian regime of transport ($(s+0.5)/(s-1)\approxeq1$) is then given by,
\begin{equation}
L_{\mathrm{s}}^{\mathrm{aeolian}}=3c_v{V_{\mathrm{s}}}V_{\mathrm{rs}}\mathrm{F}\mathrm{K}\cdot{[\mu g]}^{-1}. \label{LVfinal2aeol}
\end{equation}
%KANN MAN FOLGENDEN PARAGRAPHEN NICHT WEGLASSEN?
We show in Section IV of the Suppl.~Mat.~\cite{Suppl_Mat} that Eq.~(\ref{LVfinal2aeol}) can be approximated by the simpler form of $L_{\mathrm{s}}^{\mathrm{aeolian}} \approx 3c_v{V_{\mathrm{s}}^2}\cdot{[\mu g]}^{-1}$ in the limit of large $u_{\ast}/u_{\mathrm{t}}$.
%We note that $K$ can be approximated as its limit for large dimensionless shear velocities for Earth and Mars conditions with $d \geq 200\mu$m, resulting in the simple expression $L_{\mathrm{s}}^{\mathrm{aeolian}} \approx 3c_v{V_{\mathrm{s}}^2}\cdot{[\mu g]}^{-1}$ \cite{Suppl_Mat}.
%($(u_{\ast}/u_{\mathrm{t}})^2 \gg 1$), giving $K\approx V_{\mathrm{s}}/(FV_{\mathrm{rs}})$. This approximation has an error typically smaller than $30\%$ for Earth and Mars conditions with $d \geq 200\mu$m \cite{Suppl_Mat}. The benefit of this approximation is that it results in a simple expression for $L_\mathrm{s}$ for aeolian transport, $L_{\mathrm{s}}^{\mathrm{aeolian}} \approx 3c_v{V_{\mathrm{s}}^2}\cdot{[\mu g]}^{-1}$.

Therefore, from our general expression for $L_{\mathrm{s}}$ (Eq.~(\ref{LVfinal2})) we obtain two expressions --- Eqs.~(\ref{LVfinal2subq}) and (\ref{LVfinal2aeol}) --- which can be used to predict $L_{\mathrm{s}}$ in the subaqueous and aeolian transport regimes, respectively. Both use only two parameters, namely $\mu$ and $c_v$, which
%Indeed, while the classical scaling $L_\mathrm{s}=2sd$ \cite{Andreotti_et_al_2010,Fourriere_et_al_2010} uses a empirical pre-factor `2' which was fitted to aeolian measurements of $L_{\mathrm{s}}$ (Fig.~\ref{fig:aeolian}), the physical parameters $\mu$ and $c_v$ 
are estimated from independent measurements. Specifically, ${\mu}$ is estimated from measurements of $M_{\mathrm{s}}$ and $Q_{\mathrm{s}}$ for different values of $u_{\ast}$ in air and under water, while $c_v$ is estimated from measurements of the particle velocity distribution \cite{Suppl_Mat,Lajeunesse_et_al_2010,Creyssels_et_al_2009}. From these experimental data, we obtain $\mu \approx 1.0$ ($0.5$) and $c_v \approx 1.3$ ($1.7$) for the aeolian (subaqueous) regime.

Both Eqs.~(\ref{LVfinal2subq}) and (\ref{LVfinal2aeol}) are consistent with the behavior of $L_{\mathrm{s}}$ with $u_{\ast}$ observed in experiments. Indeed, $L_{\mathrm{s}}$ mainly depends on $u_{\ast}$ via the average particle velocity, $V_{\mathrm{s}}$. For subaqueous transport, in which $V_{\mathrm{s}}$ is a linear function of $u_{\ast}$, $L_{\mathrm{s}}$ varies linearly with $V_{\mathrm{s}}$ and thus with $u_{\ast}$, which is consistent with experiments \cite{Franklin_and_Charru_2011}. In contrast, $V_{\mathrm{s}}$ depends only weakly on $u_{\ast}$ for aeolian transport \cite{Kok_et_al_2012,Duran_et_al_2012}. Consequently, $L_{\mathrm{s}}$ is only weakly dependent on $u_{\ast}$ in this regime, which is also consistent with experiments \cite{Andreotti_et_al_2010}. In fact, when neglecting this weak dependence on $u_{\ast}$, Eq.~(\ref{LVfinal2aeol}) 
reduces to $L_{\mathrm{s}}\propto sd$ \cite{Andreotti_et_al_2010, Fourriere_et_al_2010} in the limit of large particle Reynolds numbers $\sqrt{sgd^3}/\nu$ for which $V_{\mathrm{s}}\propto\sqrt{sgd}$ \cite{Duran_et_al_2012}. Moreover, we estimate the average particle velocity $V_{\mathrm{s}}$ as a function of $u_{\ast}/u_\mathrm{t}$ using well-established theoretical expressions which were validated against experiments of sediment transport in the aeolian or in the subaqueous regime. Specifically, we use the model of Ref.~\cite{Paehtz_et_al_2012} for obtaining $V_{\mathrm{s}}(u_{\ast}/u_{{\ast}{\mathrm{t}}})$ in the aeolian regime and the model of Ref.~\cite{Lajeunesse_et_al_2010} for the subaqueous regime \cite{Suppl_Mat}.
\begin{figure}[!htpb]
 \begin{center}
  \includegraphics[width=1.00\columnwidth]{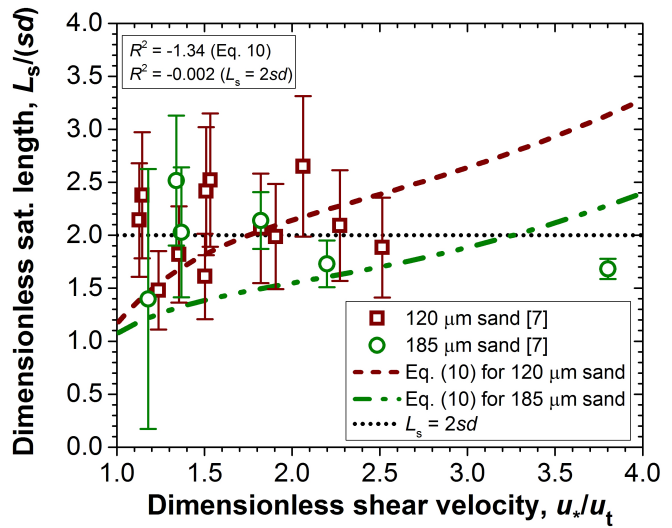}
 \caption{Dimensionless saturation length, $L_{\mathrm{s}}/(sd)$, versus 
%dimensionless shear velocity, $u_{\ast}/u_{\mathrm{t}}$, 
$u_{\ast}/u_{\mathrm{t}}$
for aeolian transport under terrestrial conditions. Brown squares denote estimates of $L_{\mathrm{s}}$ from wind-tunnel measurements ($d=\,120{\mu}$m), while the error bars are due to uncertainties in the measurements of the sediment flux \cite{Andreotti_et_al_2010}. Green circles denote $L_{\mathrm{s}}$ obtained from the wavelength of elementary dunes on top of large barchans ($d=185\,{\mu}$m), whereas the 
%associated 
error bars contain uncertainties in the dune size \cite{Andreotti_et_al_2010} (potential systematic uncertainties \cite{Suppl_Mat} are not included). The coloured lines represent predicted values of $L_{\mathrm{s}}$ using Eq.~(\ref{LVfinal2}) for the corresponding experimental conditions ($\rho_{\mathrm{p}}=2650$\,kg$/$m$^3$, $\rho_{\mathrm{f}}=1.174$\,kg$/$m$^3$ and $\nu=1.59\times10^{-5}$\,m$^2/$s). The dotted horizontal line indicates the prediction of $L_{\mathrm{s}}$ using 
%the scaling 
$L_{\mathrm{s}}= 2sd$ \cite{Andreotti_et_al_2010,Fourriere_et_al_2010}. The upper legend displays the corresponding values of the coefficient of determination, $R^2=1-\frac{\sum_i\left(L_{\mathrm{s}i}^\mathrm{measured}-L_{\mathrm{s}i}^\mathrm{predicted}\right)^2}{\sum_i\left(L_{\mathrm{s}i}^\mathrm{measured}-L_{\mathrm{s}}^{\mathrm{mean}}\right)^2}$, which is a measure of a theory's ability to capture variation in data, with $R^2$ = 1 corresponding to a perfect fit).}
 \label{fig:aeolian}
 \end{center}
\end{figure}
\begin{figure}[!htpb]
 \begin{center}
  \includegraphics[width=1.00\columnwidth]{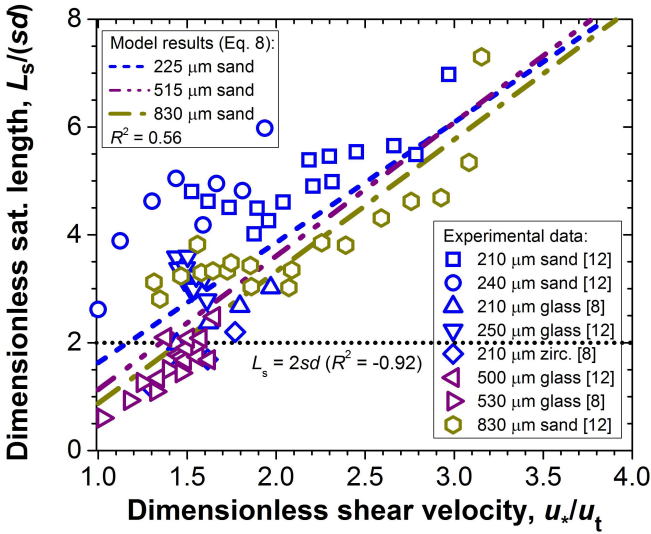}
 \caption{$L_{\mathrm{s}}/(sd)$ versus $u_{\ast}/u_{\mathrm{t}}$ for subaqueous transport. Symbols denote estimates of $L_{\mathrm{s}}$ from the wavelength of elementary dunes \cite{Fourriere_et_al_2010} and from the minimal cross-stream width of subaqueous barchans, $W\! \approx\! 12\,L_{\mathrm{s}}$ \cite{Franklin_and_Charru_2011}. The coloured lines denote predicted values of $L_{\mathrm{s}}$ using Eq.~(\ref{LVfinal2}) for subaqueous transport of sand (${\rho}_{\mathrm{p}}\!=\!2650\,$kg$/$m$^3$, $\rho_{\mathrm{f}}\!=\!10^3\,$kg$/$m$^3$ and $\nu\!=\!10^{-6}\,$m$^2/$s), with grain sizes roughly matching those used in the experiments. The dotted horizontal line indicates the prediction of $L_{\mathrm{s}}$ using the scaling $L_{\mathrm{s}}= 2sd$ \cite{Andreotti_et_al_2010,Fourriere_et_al_2010}. The values of $R^2$ (coefficient of determination) for both expressions are also shown.
} 
 \label{fig:subaqueous}
\end{center}
\end{figure}

The squares in Fig.~\ref{fig:aeolian} denote wind tunnel measurements of $L_{\mathrm{s}}$ for different values of $u_{\ast}$. These data were obtained by fitting Eq.~(\ref{Ls}) to the downstream evolution of the sediment flux, $Q(x)$, close to equilibrium \cite{Andreotti_et_al_2010}. Further estimates of $L_{\mathrm{s}}$ for aeolian transport under terrestrial conditions have been obtained from the wavelength ($\lambda$) of elementary dunes on top of large barchans \cite{Andreotti_et_al_2010,Suppl_Mat}. These estimates correspond to the circles in Fig.~\ref{fig:aeolian}, whereas the coloured lines in this figure denote $L_{\mathrm{s}}/(sd)$ versus $u_{\ast}/u_{\mathrm{t}}$ predicted by Eq.~(\ref{LVfinal2aeol}). As we can see in Fig.~\ref{fig:aeolian}, in spite of the scatter in the data, Eq.~(\ref{LVfinal2aeol}) yields reasonable agreement with the experimental data {\em{without requiring any fitting to these data}}. In contrast, the scaling $L_\mathrm{s}=2sd$ \cite{Andreotti_et_al_2010,Fourriere_et_al_2010} was obtained from a fit to the data displayed in Fig.~\ref{fig:aeolian}. Moreover, Fig.~\ref{fig:subaqueous} shows values of $L_{\mathrm{s}}$ estimated from experiments on subaqueous transport under different shear velocities (symbols). These estimates were obtained from measurements of $\lambda$ \cite{Fourriere_et_al_2010,Langlois_and_Valance_2007} and from the minimal cross-stream width, $W \approx 12L_{\mathrm{s}}$ \cite{Parteli_et_al_2007}, of barchans in a water flume \cite{Franklin_and_Charru_2011}. The coloured lines show the behavior of $L_{\mathrm{s}}$ with $u_{\ast}$ as predicted from Eq.~(\ref{LVfinal2subq}) for subaqueous sand transport. We note that Eq.~(\ref{LVfinal2subq}) is the first expression for $L_\mathrm{s}$ that shows good agreement with measurements of $L_{\mathrm{s}}$ under water. Indeed, the scaling relation $L_\mathrm{s}=2sd$
%\cite{Andreotti_et_al_2010,Fourriere_et_al_2010}, which was developed for the aeolian regime, 
does not capture the increasing trend of $L_{\mathrm{s}}$ with $u_\ast/u_\mathrm{t}$ evident from the experimental data. 

\begin{figure}[!ht]
\begin{center} 
\includegraphics[width=1.00\columnwidth]{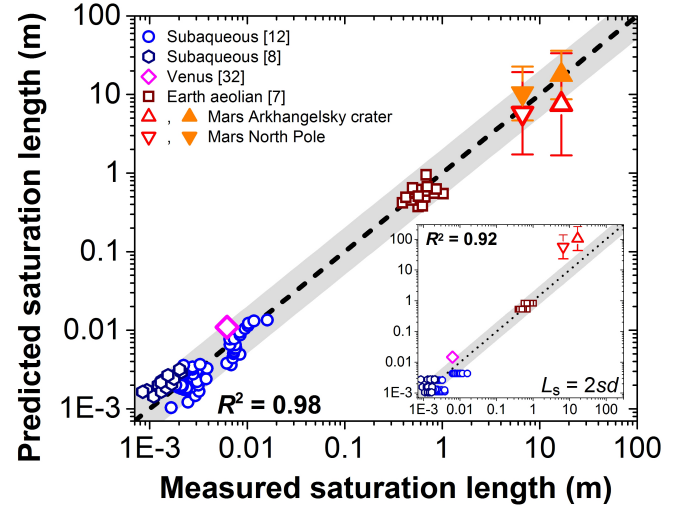}
\caption{Comparison of measured and predicted values of $L_{\mathrm{s}}$ for various environments; the grey shading denotes agreement between measurements and Eqs.~(\ref{LVfinal2aeol}) and (\ref{LVfinal2subq}) within a factor of two. Aeolian and subaqueous data were obtained as described in Figs.~1 and 2. For Venus, $L_{\mathrm{s}}$ was estimated from the wavelength of elementary dunes \cite{Suppl_Mat}. For Mars, $L_{\mathrm{s}}$ was derived from estimates of the minimal size of barchans in two Martian dune fields \cite{Suppl_Mat}, and plotted against the predicted $L_{\mathrm{s}}$ for a range of plausible dune particle sizes ($d = 100 - 600\,{\mu}$m \cite{Bourke_et_al_2010}) and for two separate estimates of $u_{{\ast}{\mathrm{typ}}}$ (see text). 
%Specifically, the 
The error bars denote the range in predicted $L_{\mathrm{s}}$ arising from the range in $d$ \cite{Bourke_et_al_2010}, and the symbols denote the geometric mean. The filled orange symbols use $u_{{\ast}{\mathrm{typ}}} = u_{\mathrm{ft}}$ \cite{Parteli_et_al_2007,Sullivan_et_al_2005}, while the open red symbols use the $u_{{\ast}{\mathrm{typ}}}$ calculated from Viking 2 wind speed measurements \cite{Suppl_Mat}. The inset shows measured and predicted values of $L_{\mathrm{s}}$ for the same conditions as in the main plot, but using 
%the scaling 
$L_{\mathrm{s}} = 2sd$ \cite{Andreotti_et_al_2010,Fourriere_et_al_2010}.
%In each plot, $R^2$ gives the coefficient of determination.
Values of $R^2$ (coefficient of determination) in each plot were calculated in log10-space such that each data point was weighted equally.}
\label{fig:planetary} 
\end{center} 
\end{figure}

An excellent laboratory for further testing our model is the surface of Mars, where the ratio of grain to fluid density ($s$) is about two orders of magnitude larger than on Earth. We estimate the Martian $L_{\mathrm{s}}$ from reported values of the minimal crosswind width $W$ of barchans at 
%wo dune fields on Mars: one at 
Arkhangelsky crater in the southern highlands and 
at a dune field
%the other 
near the north pole \cite{Parteli_et_al_2007,Suppl_Mat}. However, using Eq.~(\ref{LVfinal2aeol}) to predict $L_{\mathrm{s}}$ on Mars is difficult because both the grain size $d$ and the typical shear velocity $u_{{\ast}{\mathrm{typ}}}$ for which the dunes were formed are poorly known. Indeed, we need to know both quantities to calculate $V_{\mathrm{s}}$ \cite{Paehtz_et_al_2012}. We thus predict the Martian $L_{\mathrm{s}}$ using a range of plausible values of $d$ and $u_{{\ast}{\mathrm{typ}}}$. Specifically, we assume $d$ to lie in the broad range of $100-600\,{\mu}$m based on recent studies \cite{Bourke_et_al_2010}. Estimating $u_{{\ast}{\mathrm{typ}}}$ on Mars is also difficult, both because of the scarcity of wind speed measurements \cite{Sullivan_et_al_2000}, and because the threshold $u_{\ast}$ required to initiate transport ($u_{\mathrm{ft}}$) likely exceeds $u_{\mathrm{t}}$ by up to a factor of~$\sim 10$ \cite{Kok_et_al_2012,Paehtz_et_al_2012,Kok_2010}. We therefore calculate $L_{\mathrm{s}}$ for two separate estimates of $u_{{\ast}{\mathrm{typ}}}$: the first using $u_{{\ast}{\mathrm{typ}}}=u_{\mathrm{ft}}$, consistent with previous studies \cite{Parteli_et_al_2007,Sullivan_et_al_2005}, and the second calculating $u_{{\ast}{\mathrm{typ}}}$ based on the wind speed probability distribution measured at the Viking 2 landing site \cite{Suppl_Mat}, which results in an estimate of $u_{{\ast}{\mathrm{typ}}}$ closer to $u_{\mathrm{t}}$. Fig.~3 shows that the values of $L_{\mathrm{s}}$ predicted with either of these estimates are consistent with those estimated from the minimal barchan width. This good agreement suggests that the previously noted overestimation of the minimal size of Martian dunes \cite{Kroy_et_al_2005} is largely resolved by accounting for the low Martian value of $u_{\mathrm{t}}/u_{\mathrm{ft}}$ \cite{Paehtz_et_al_2012} and the proportionally lower value of the particle speed $V_{\mathrm{s}}$, as hypothesized in Ref.~\cite{Kok_2010}. 
Indeed, the scaling $L_{\mathrm{s}} = 2sd$ (inset of Fig.~\ref{fig:planetary}) requires $d \approx29\mu$m and $d \approx40\mu$m to be consistent with $L_{\mathrm{s}}$ for the north polar and Arkhangelsky dune fields, respectively. However, such particles
%, which are smaller than the fines within granule ripples on Mars \cite{Sullivan_et_al_2005} ($87\pm15\mu$m), 
are most likely transported as suspended load 
on Mars
%in the Martian atmosphere 
\cite{Sullivan_et_al_2005}, as they are on Earth \cite{Shao_2008}.

Finally, Fig.~\ref{fig:planetary} also compares Eq.~(\ref{LVfinal2subq}) to measurements of $L_{\mathrm{s}}$ for Venusian transport, which have been estimated from the wavelength of elementary dunes produced in a wind-tunnel mimicking the Venusian atmosphere \cite{Marshall_and_Greeley_1992}.

In conclusion, Eq.~(\ref{LVfinal2})
%our expression for the saturation length ($L_{\mathrm{s}}$) of sediment flux (Eq.~(\ref{LVfinal2})) 
is the first 
expression
capable of quantitatively reproducing measurements of 
the saturation length
$L_{\mathrm{s}}$ under different flow conditions in both air and under water, and is in agreement with measurements over at least 5 orders of magnitude of variation in the sediment to fluid density ratio. 
%The capability of our expression to reproduce measurements of $L_{\mathrm{s}}$ in a wide range of physical environments without requiring fitting to any of the experimental data attests for the significant advance in the understanding of the physics of sediment transport achieved through our model for the saturation length. 
The future application of 
%our 
this
expression thus has the potential to provide important contributions to calculate sediment transport, the response of saltation-driven wind erosion and dust aerosol emission to turbulent wind fluctuations, and the dynamics of sediment-composed landscapes under water, on Earth's surface and on other planetary bodies. 

We acknowledge support from grants NSFC 41350110226, NSFC 41376095, ETH-10-09-2, NSF AGS 1137716, and DFG through the Cluster of Excellence ``Engineering of Advanced Materials''. We thank Miller Mendoza and Robert Sullivan for discussions, and Jeffery Hollingsworth for providing us with the pressure and temperature at the Martian dune fields.

%\bibliography{model} 

\end{document}